\documentclass{aa}

\usepackage{graphicx}
\usepackage{dcolumn}
\usepackage{bm}
\usepackage{txfonts}
\usepackage{natbib}
\usepackage{amsmath}    
\usepackage{graphicx}   
\usepackage{verbatim}   
\usepackage{color}      
\usepackage{subfigure}  
\usepackage{gensymb} 
\usepackage{wasysym} 
\usepackage{hyperref}   
\bibpunct{(}{)}{;}{a}{}{,} 
\usepackage{graphicx}
\usepackage{nth}
\newcommand{\kms}{\mbox{${\rm km\,s}^{-1}$}}

\newcommand{\Msolar}{\mbox{${M}_{\astrosun}$}}
\newcommand{\Rsolar}{\mbox{${R}_{\astrosun}$}}

\newcommand{\rhosun}{\mbox{$\rho_{\astrosun}$}}
\newcommand{\Rjup}{\mbox{${R}_{J}$}}

\newcommand{\vsini}{\mbox{$v \sin i_{\ast}$}}

\newcommand\T{\rule{0pt}{2.2ex}}

\newcommand{\NaI}{\ion{Na}{i}}

\begin{document}

\title{Hot Exoplanet Atmospheres Resolved 
with Transit Spectroscopy (HEARTS)\thanks{Based on observations made at ESO 3.6~m telescope (La Silla, Chile) under ESO programmes 090.C-0540 and 100.C-0750.}}
\subtitle{II. A broadened sodium feature on the ultra-hot giant WASP-76b}
\author{J.~V.~Seidel\inst{1} 
\and D.~Ehrenreich\inst{1}
\and A.~Wyttenbach\inst{2}
\and R.~Allart\inst{1}
\and M.~Lendl\inst{3,1}
\and L.~Pino\inst{4}
\and V.~Bourrier\inst{1}
\and H.~M.~Cegla\inst{1,5}
\and C.~Lovis\inst{1}
\and D.~Barrado\inst{6}
\and D.~Bayliss\inst{7}
\and N.~Astudillo-Defru\inst{8}
\and A.~Deline\inst{1}
\and C.~Fisher\inst{9}
\and K.~Heng\inst{9}
\and R.~Joseph\inst{10}
\and B.~Lavie\inst{1,9}
\and C.~Melo\inst{11}
\and F.~Pepe\inst{1}
\and D.~S\'egransan\inst{1}
\and S.~Udry\inst{1}
}

\institute{Observatoire astronomique de l'Universit\'e de Gen\`eve, chemin des Maillettes 51, 1290 Versoix, Switzerland
\and Leiden Observatory, Leiden University, Postbus 9513, 2300 RA Leiden, The Netherlands
\and Space Research Institute, Austrian Academy of Sciences, Schmiedlstr. 6, 8042, Graz, Austria
\and Anton Pannekoek Institute for Astronomy, University of Amsterdam, Science Park 904, 1098 XH Amsterdam, The Netherlands
\and CHEOPS Fellow, SNSF NCCR-PlanetS
\and Depto. Astrof\'isica, Centro de Astrobiolog\'ia (INTA-CSIC), ESAC campus, Camino Bajo del Castillo s/n, 28692, Villanueva de la Ca\~nada, Spain
\and Department of Physics, University of Warwick, Gibbet Hill Rd., Coventry, CV4 7AL, UK
\and Universidad de Concepci\'on, Departamento de Astronom\'ia, Casilla 160-C, Concepci\'on, Chile
\and University of Bern, Center for Space and Habitability, Gesellschaftsstrasse 6, CH-3012, Bern, Switzerland
\and Laboratoire d'Astrophysique, \'Ecole Polytechnique F\'ed\'erale de Lausanne (EPFL), Observatoire de Sauverny, 1290 Versoix, Switzerland
\and European Southern Observatory, Alonso de C\'ordova 3107, Vitacura, Regi\'on Metropolitana, Chile
}

\abstract{High-resolution optical spectroscopy is a powerful tool to characterise exoplanetary atmospheres from the ground. The sodium D lines, with their large cross sections, are especially suited to study the upper layers of atmospheres in this context. We report on the results from HEARTS, a spectroscopic survey of exoplanet atmospheres, performing a comparative study of hot gas giants to determine the effects of stellar irradiation. In this second installation of the series, we highlight the detection of neutral sodium on the ultra-hot giant WASP-76b.  We observed three transits of the planet using the HARPS high-resolution spectrograph at the ESO 3.6m telescope and collected 175 spectra of WASP-76. We repeatedly detect the absorption signature of neutral sodium in the planet atmosphere ($0.371\pm0.034\%$; $10.75~\sigma$ in a $0.75$~\r{A} passband). The sodium lines have a Gaussian profile with full width at half maximum (FWHM) of $27.6\pm2.8~\kms$. This is significantly broader than the line spread function of HARPS ($2.7~\kms$). We surmise that the observed broadening could trace the super-rotation in the upper atmosphere of this ultra-hot gas giant.}
\keywords{Planetary Systems -- Planets and satellites: atmospheres, individual: WASP-76b -- Techniques: spectroscopic -- Instrumentation: spectrographs -- Methods: observational}
\maketitle

\section{Introduction}

The field of exoplanetary research has moved firmly into the era of atmospheric characterisation. 
To learn more about the planet atmospheric composition and conditions, the light arriving from the host star during the transit is commonly used to extract the spectrum generated by the atmosphere of the planet. 
Based on theoretical work describing transmission spectroscopy and its potential \citep{Ma99,Se00,Br01}, the first attempts to detect reflected light from the day side \citep{Ch00} or the absorption of light through limb transmission \citep{Mo01} were unsuccessful. \\

It took until 2002 to observe the first exoplanetary atmosphere on HD~209458b via the detection of atmospheric neutral sodium (\NaI) at 589 nm \citep{Ch02}. These milestone measurements were performed with the Space Telescope Imaging Spectrograph (STIS) on board the 2.4-m Hubble Space Telescope (HST). The spectral resolution of STIS is  $\mathcal{R} \equiv \lambda/\Delta \lambda \sim 5500$ ( $\sim 55$~\kms). Due to terrestrial atmospheric variation and telluric line contamination, ground based observations were unable to reproduce the success of space-based missions until six years later when \cite{Re08} and \cite{Sn08} detected atmospheric sodium in HD~189733b and HD~209458b, respectively. 

Neutral sodium is a sensitive probe for the higher atmosphere of exoplanets. Strong signatures in transit spectroscopy arise in the doublet resonant lines (\ion{Na}{i} D$_1$ and D$_2$) at 589~nm even for small amounts of sodium in the atmosphere. Given these unique features, ground-based observations have led to a wealth of sodium detections in hot exoplanet atmospheres  since then  \citep{Wo11,Je11,Zh12,Mu14,Bu15,Wy15, Ca17,Kh17,Si16,Ni16,Wy17,Che17,Je18} using spectrographs with resolutions of $\sim 100\,000$, making sodium one of the most often detected species to date.

Atmospheric neutral sodium is not only a powerful tool to detect exoplanetary atmospheres, but also to characterise their properties. With a high cross section, the \NaI~doublet (at 589 nm, also called the Frauenhofer D line) probes the atmosphere up to high altitudes, thus constraining the temperature pressure profile \citep{Vi11a,Vi11b,He15}  and other dynamical processes up to the thermosphere \citep{Lo15,Wy15}. 

This technique has been successfully applied to HD~189733b \citep{Wy15,Pi18} with HARPS, a high resolution spectrograph stabilised in temperature and pressure at the 3.6m telescope of ESO at La Silla Observatory, Chile. \cite{Wy15} thus established the usefulness of HARPS in the search for atmospheric signatures during exoplanet transits. Based on this benchmark result the Hot Exoplanet Atmosphere Resolved with Transit Spectroscopy survey (HEARTS) was created to observe a large sample of gas giants with different masses and irradiation with HARPS. 

\begin{figure}[htb]
\resizebox{\columnwidth}{!}{\includegraphics[trim=2.5cm 1.0cm 1.0cm 1.5cm]{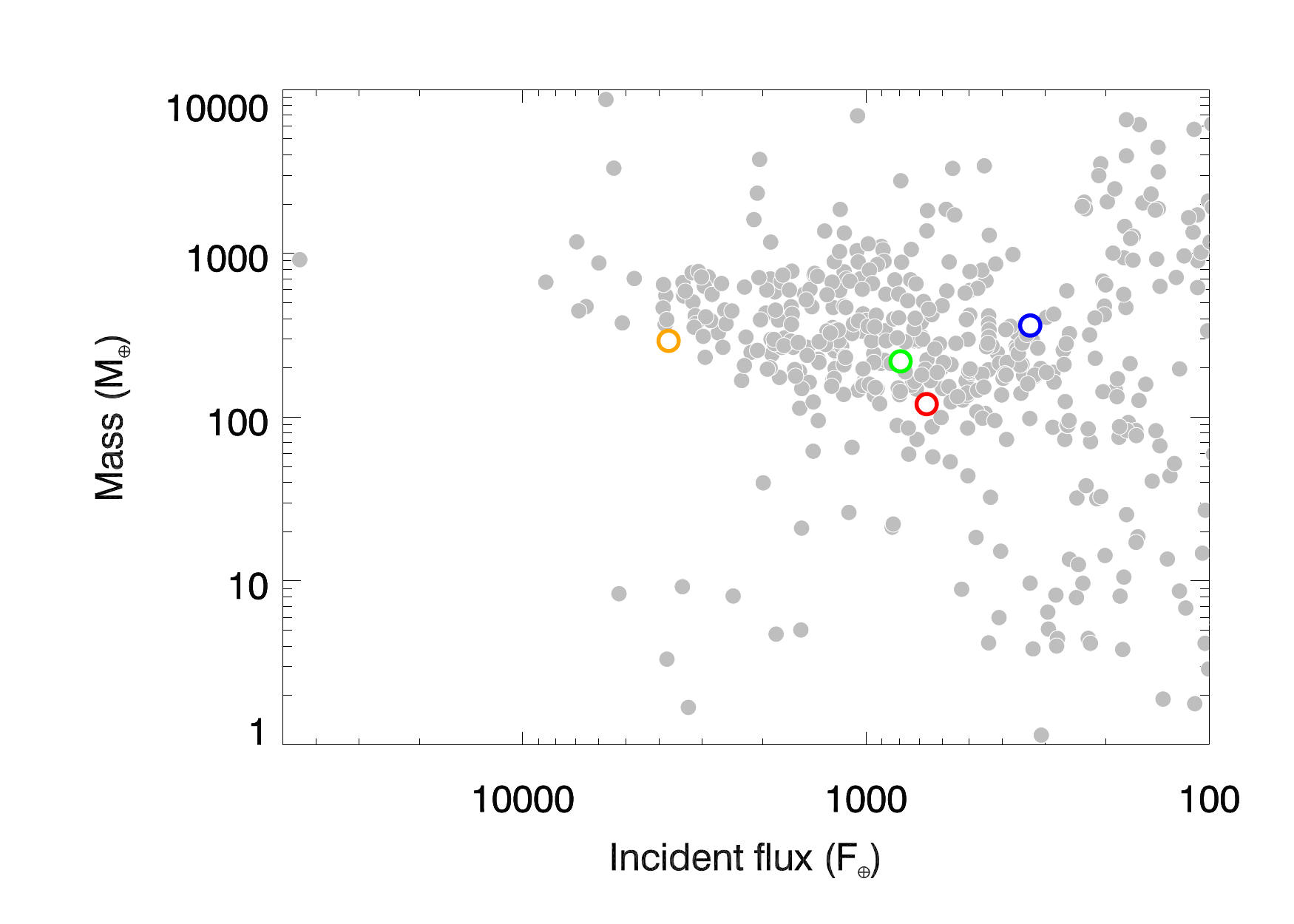}}
	\caption{Mass of the planet vs. bolometric optical and infrared incident flux in units of flux received at Earth. In grey, transiting exoplanets are shown with $V < 16$ and masses determined with a precision better than $20~\%$. WASP-76b is shown in orange. For comparison WASP-49b  \citep{Wy17}, HD~189733b \citep{Wy15} and HD~209458b (for reference) are shown in red, blue and green respectively.}
	\label{fig:massflux}
\end{figure}

The first result from this study was the detection of hot neutral sodium at high altitudes on WASP-49b \citep{Wy17}. This paper is the second installation from the HEARTS survey and reports on the detection of neutral sodium in WASP-76b, which allows us to unveil the upper atmosphere properties of the planet. WASP-76b is a gas giant of approximately one Jupiter mass, but roughly twice its radius.  The WASP-76b properties are listed in Tables \ref{table:systemoverview} and \ref{tab:para} and a comparison to similar planets of sub-Jupiter mass with high irradiation can be seen in Fig. \ref{fig:massflux}.

In the following section, we describe the observations with HARPS and their data reduction (Section \ref{sec:obs}), followed by Section \ref{sec:EulerCam} describing the simultaneous photometric observations of WASP-76 with EulerCam. Section \ref{sec:fulltrans} highlights the detection of neutral sodium from the transmission spectrum, the evolution of the signal with time and discusses the observed sodium line profile.

\begin{table}
\caption{Orbital and physical parameters of the WASP-76 system \citep{We16}. All other relevant parameters have been updated in this work (see Section \ref{sec:EulerCam}).}
\label{table:systemoverview}
\centering
\begin{tabular}{l c }
\hline
\hline
Parameter 	& Value 	\\
\hline
&\\
$V$ mag & 9.5\\
Spectral type & F7\\
$M_{\star}$ & $1.46 \pm 0.07$ $M_{\astrosun}$\\
$R_{\star}$ & $1.73 \pm 0.04$ $R_{\astrosun}$ \\
$T_{eff}$ & $6250\pm 100$ K\\
$\log  g$& $4.4 \pm 0.1$\\

   K$_1$ & $0.1193 \pm 0.0018$ km s$^{-1}$\\
   $e$ & 0 (assumed)\\
    $\omega$ &  90 (assumed)  \\
      $\gamma$ & $-1.0733 \pm 0.0002$ km s$^{-1}$\\
 M$_p$ & $0.92 \pm 0.03$ M$_{Jup}$ \\
   \hline
\end{tabular}
\end{table}

\section{Observations and data reduction}
\label{sec:obs}
\subsection{HARPS observations}
 We collected data of three transits of the hot gas giant planet WASP-76b with its host star WASP-76, a F7 star with a visual apparent magnitude of V=$9.5$. The observations were done with the HARPS (High-Accuracy Radial-velocity Planet Searcher \citep{Ma03}) spectrograph at the ESO 3.6 m telescope in La Silla Observatory, Chile. We performed observations on 2017-10-24 and 2017-11-22 as part of the HEARTS survey (ESO programme: 100.C-0750; PI: Ehrenreich) and combined them with an archival data set obtained on 2012-11-11 (ESO programme: 090.C-0540; PI: Triaud; \cite{Brown17}).

 \begin{table*}
\caption{Log of observations.}
\label{table:nightoverview}
\centering
\begin{tabular}{c c c c c c c }
\hline
\hline
&Date	&$\#$Spectra \tablefootmark{a}	&Exp. Time [s]	&Airmass	\tablefootmark{b}&Seeing	&SNR order 56 \tablefootmark{c}	\\
\hline
Night 1&2012-11-11&61 (21/40)&300&1.7-1.18-1.5&Not recorded&27.9 -  40.1\\
Night 2&2017-10-24&49 (22/27)&600,400,350&1.7-1.18-2.0&0.8-1.4&49.7 -  68.9\\
Night 3&2017-11-22&65 (25\tablefootmark{d}/40)&300&1.4-1.18-2.3&0.6-1.0&42.1 - 65.3\\
\hline
\end{tabular}
\tablefoot{\tablefoottext{a}{In parenthesis: spectra in and out-of-transit respectively.}
\tablefoottext{b}{Airmass at the beginning, centre and end of transit.}\tablefoottext{c}{Order 56 contains the sodium feature.}\tablefoottext{d}{From the 25 out-of-transit exposures, 15 exposures after the transit were disregarded due to cloud contamination. The out-of-transit baseline was built with the 10 remaining exposures before the transit.}}
\end{table*}

 In the HEARTS survey, each target is observed for multiple transits to ensure reproducibility of the recorded spectra and a high signal to noise ratio. The HARPS spectra are extracted order by order from the 2D echelle spectral images (stored in e2ds files) by the HARPS Data Reduction Pipeline (DRS v3.5) In this study, we concentrated our analysis on order number 56, which contains the sodium doublet. The spectral order $56$ samples the wavelength region from $5850.24$~\r{A} to $5916.17$~\r{A}. 
  The data is corrected for the blaze and the pixel grid is calibrated to the wavelength solution grid using the daily afternoon calibration. The wavelength are extracted from the header provided by HARPS and are given in air in the solar system barycentric rest frame. Each observation is started as early as possible with respect to the transit and ends when the star is not observable any more, which allows us to record the transit event and the best possible out-of-transit baseline before and after the transit. A good baseline is necessary to build an accurate out-of-transit master spectrum ("master out" spectrum) and thus an accurate spectrum ratio. 
A log of the observations can be found in Table \ref{table:nightoverview}). All transits were observed with one fibre on the target (fibre A) and one fibre on the sky (fibre B). This technique makes a correction for telluric contamination necessary (see Section \ref{sec:tell}). The exposure times varied from 300s to 600s depending on seeing conditions in the respective nights. The three nights resulted in a total of 175 spectra with 68 in-transit and 107 out-of-transit. In the third night, the out-of-transit exposures after the transit (15 spectra) show a lower signal-to-noise ratio due to cirrus clouds and were discarded from the analysis. This leaves ten out-of-transit exposures taken before the ingress to build the out-of-transit baseline. The in-transit exposures are defined as spectra recorded fully or partially during the transit of the planet (i.e. between the first and fourth contacts), the out-of-transit exposures are recorded before and after the transit event with $\sim 1$ hour before and $\sim 2\!-\!3$ hours after the transit.

\subsection{Telluric correction with {\tt molecfit}}
\label{sec:tell}

\begin{figure}
\resizebox{\columnwidth}{!}{\includegraphics[trim=4.0cm 9cm 3.5cm 9.5cm]{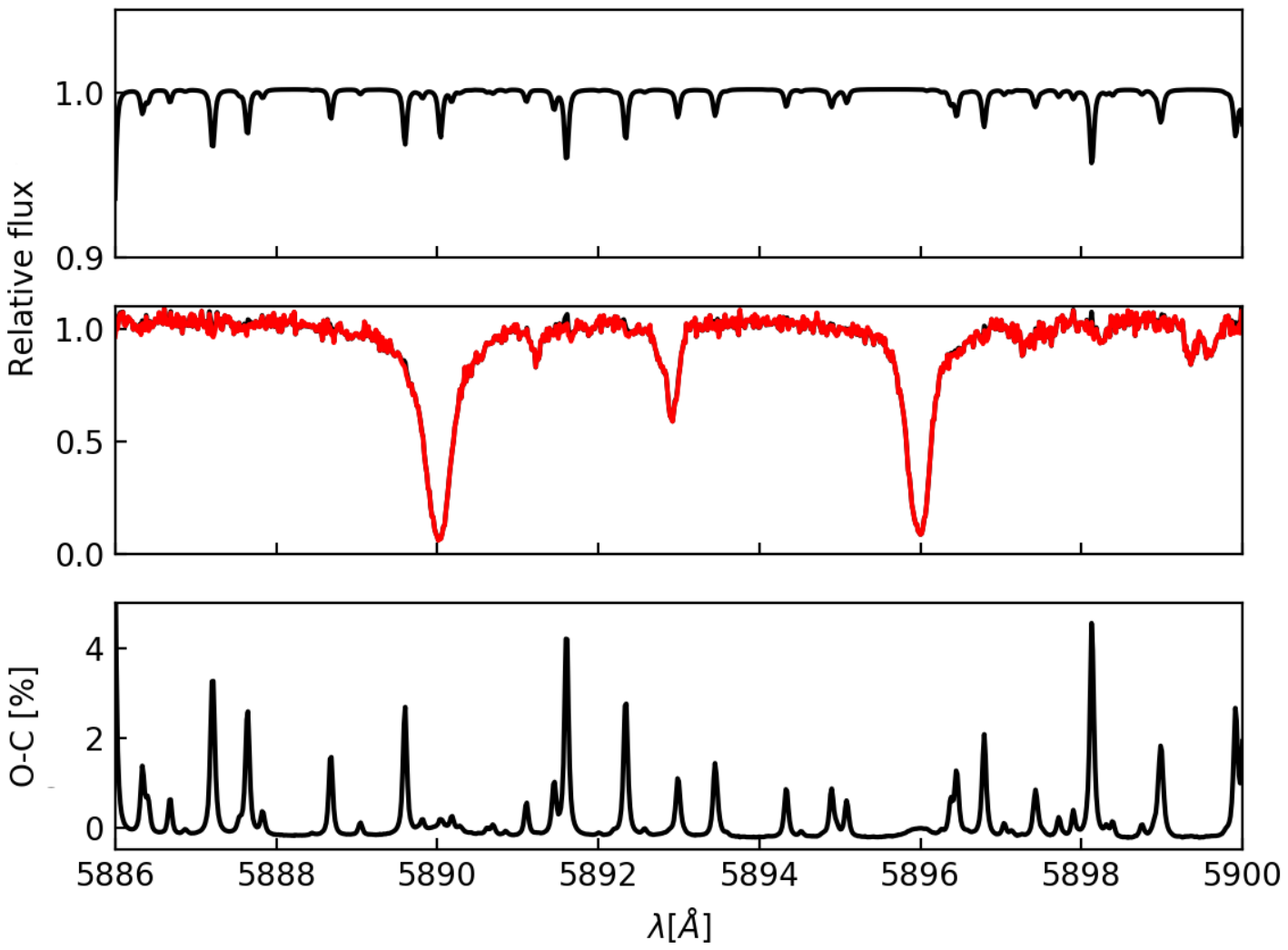}}
	\caption{The telluric correction is highlighted for an exposure out-of-transit during the first observation night on 2012-11-11. The upper panel shows the applied model of telluric lines. The middle panel shows the uncorrected flux in red and the corrected flux in black, which cannot be distinguished. Note the different scales for the two panels. The lower panel shows the difference between the corrected and uncorrected spectra for better visualisation. The impact of tellurics in the wavelength range of the sodium doublet's cores is negligible. Nonetheless a good telluric correction is crucial to clean the continuum baseline used for normalisation and where more telluric lines are situated.}
\label{fig:tellover}
\end{figure}

Figure \ref{fig:tellover} shows the influence of telluric lines in the optical ground-based observations of the HARPS spectral order containing sodium. As can be seen in Fig. \ref{fig:tellover}, the optimal telluric correction does not impact the exposures in the wavelength range of the sodium doublet. However, telluric lines can influence the normalisation processes during the analysis since the lines appear stronger in the continuum. Furthermore we wish to use this pipeline for future observations with HARPS, HARPS-N and ESPRESSO which are at different locations. A stable correction for tellurics becomes more influencial when dealing with HARPS-N (Roque de los Muchachos, La Palma) or during different seasons. To perform the correction we used {\tt molecfit} version 1.5.1. \citep{Sm15, Ka15}, a tool to correct telluric features in ground-based observations provided by ESO. This approach has been used for the first time on a HARPS spectrum by \cite{Al17}. The {\tt molecfit} software computes a high resolution ($\lambda/\Delta \lambda \sim 4,000,000$) telluric spectrum with a line-by-line radiative transfer model (LBLRTM). 

The LBLRTM model uses an atmospheric profile as input to calculate the spectrum that depends on the temperature, pressure, humidity and abundance of the molecular species by height for a specific location and time with known airmass. To generate the input profile an atmospheric standard profile is merged with a Global Data Assimilation System (GDAS) profile. Atmospheric standard profiles provide pressure, temperature and abundance for up to tens of molecular species for different altitudes at specific latitudes and were computed by the Reference Forward Model \citep{Re01}. Additionally the GDAS profiles, which are provided by the National Oceanic and Atmospheric Administration (NOAA), contain meteorological information such as pressure, temperature and relative humidity as a function of height for specific locations and are updated every three hours. 

The GDAS grid of pressure, temperature, abundance and humidity with 100 to 150 layers in height is used to calculate the model spectrum. The model spectrum is adjusted to the observed spectrum by changing the continuum, the wavelength calibration and the instrument resolution. The only element creating telluric lines in the spectral order of sodium is H$_2$O, which allows us to omit O$_2$ and all other telluric elements from the calculation. 

\subsubsection{Application of {\tt molecfit} to WASP-76b e2ds spectra}

The e2ds pipeline of the HARPS spectrograph provides the spectra in the observer rest frame with the wavelength calibration adapted to air. Before using {\tt molecfit} on the observed spectra we correct for the blaze in each night. The telluric features are not prominent in the spectral order of sodium and the modelling regions have to be selected meticulously by hand to not only avoid stellar features, but also to find telluric lines above the noise level. The fitting process aims at adjusting the continuum, the wavelength calibration and the instrumental resolution for the telluric band of each spectrum. The convergence criteria (Levenberg-Marquardt $\chi^2$ and parameter convergence criterion)  were set to $10^{-9}$, the continuum adjusted with a first degree polynomial and the wavelength calibrated with a Chebyschev third degree polynomial (see \cite{Al17}). The instrumental profile is set to have a Gaussian form with a FWHM of 3.5 pixel. For a detailed description of all {\tt molecfit} free parameters, see \cite{Sm15}. The specific values for all {\tt molecfit} parameters used herein can be found in Table \ref{table:molecfit}.

\begin{table}
\caption{Initial parameters for {\tt molecfit} in each night based on \cite{Al17}.}
\label{table:molecfit}
\centering
\begin{tabular}{p{1.5cm} p{1cm} p{5cm} }
\hline
\hline
Initial   Parameters	& Values & Commentary 	\\
\hline
ftol & $10^{-9}$ &  $\chi^2$  convergence criterion\\
xtol & $10^{-9}$& parameter convergence criterion\\
molecules & H$_2$O& for the spectral order of sodium\\
n$_{\mathrm{cont}}$ & 1 & polynomial degree for the continuum\\
a$_0$ & $1.2,3,2$ \tablefootmark{1}& constant offset of the continuum in $10^
3$\\
calib. $\lambda$& air & type of wavelength calibration\\
n$_{\lambda}$ &$3$ & Chebyschev degree of wavelength calibration\\
b$_0$ & $0$ & constant offset for wavelength calibration\\
$\omega_{\mathrm{gaussian}}$ &$3.5$ & FWHM in pixel\\
kernel size & $15$ & \\
pixel scale & $0.16$& in arcsec/pixel\\
slit width &$1.0$& in arcsec\\
MIPAS profile &equ.atm& equatorial profile\\
atmospheric profile &$0$& natural profile\\
PWV & $-1$ & no value taken into account\\
\hline
\end{tabular}
\tablefoot{\tablefoottext{1}{The values were applied respectively to the nights 2012-11-11, 2017-10-24 and 2017-11-22}}
\end{table}

\subsubsection{Assessment of the telluric correction}

\begin{figure*}[ht]
\resizebox{\textwidth}{!}{\includegraphics[trim=1.0cm 4.5cm 1.0cm 4.5cm]{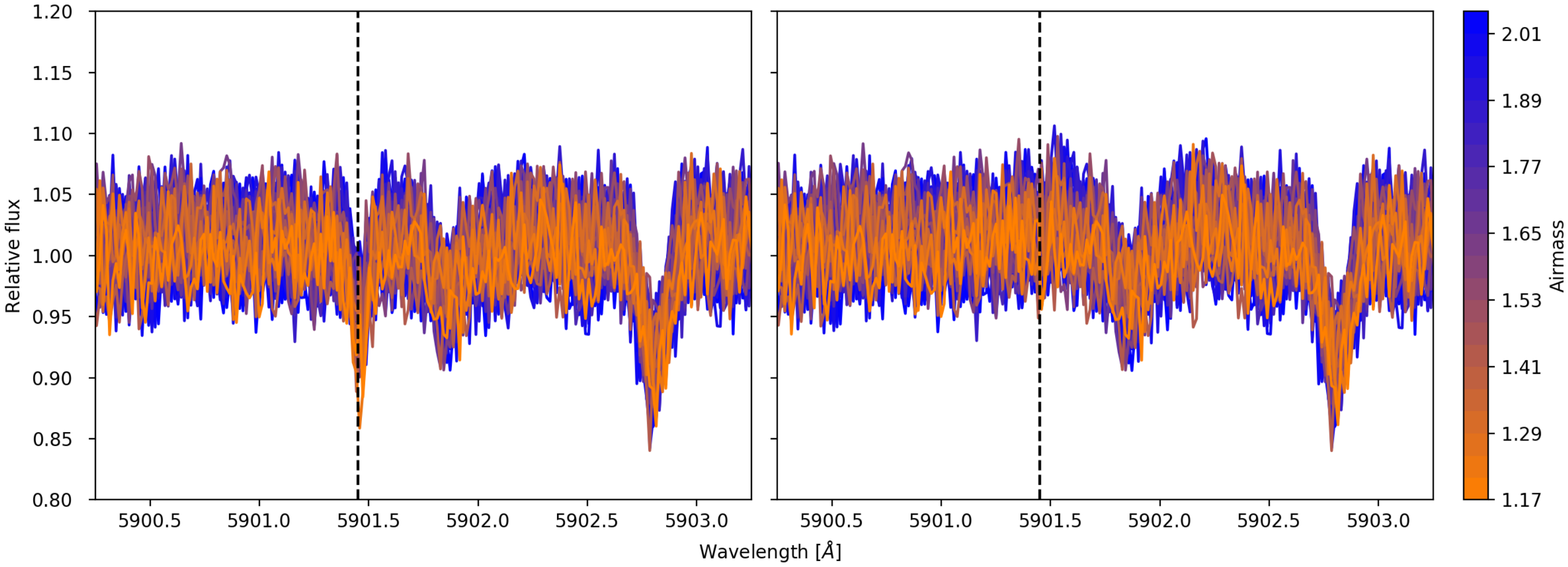}}
	\caption{All exposures for the second night (2017-10-24) plotted before (left) and after (right) telluric correction coloured by airmass. The strong telluric line (line center indicated by the black dashed line) merged with a stellar line is corrected down to the noise level for high and low airmasses. This window is illustrative of the corrections performed in all three observation nights.}
	\label{fig:tellcorr}
\end{figure*}

The {\tt molecfit} output file contains all the fitted parameters as well as the uncorrected spectrum (the input spectrum), the modelled telluric spectrum and the corrected spectrum.  In Fig. \ref{fig:tellcorr} all the corrected spectra are plotted for one night with the airmass indicated via a colour spectrum. As we can see by comparing the left and right panel of Fig. \ref{fig:tellcorr}, the telluric lines are corrected down to the noise level for all airmasses (including the lines merged with stellar lines). We checked that the telluric line highlighted in Fig. \ref{fig:tellcorr} is indeed a telluric line by comparison with the master-out spectrum for the night. Additionally, sodium emission lines can be present in the Earth's atmosphere, an effect {\tt molecfit} does not correct for. This effect can be estimated from our data taken with fiber B on the sky. Building master-out spectra for all nights from the fiber B data reveals no emission lines of sodium beyond the noise level. Sodium emission from Earth's atmosphere has therefore no impact on our findings.

\section{Simultaneous photometry with EulerCam}
\label{sec:EulerCam}
\begin{figure}
 \centering
 \label{fig:phot}
\resizebox{\columnwidth}{!}{\includegraphics[trim=0.5cm 0.5cm 0cm 0cm]{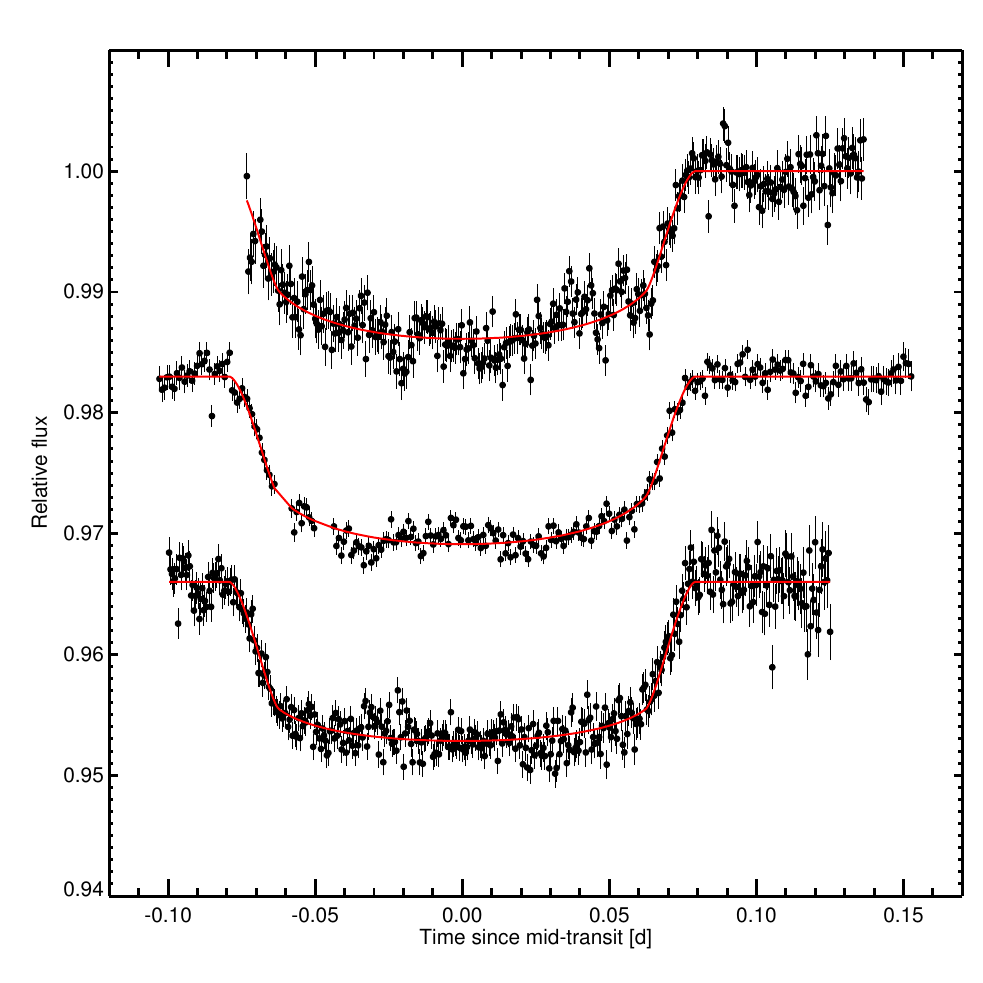}}
 \caption{The three lightcurves obtained by EulerCam (top to bottom: 2012-11-11,2017-10-24 and 2017-11-22) offset by 0.02 for visibility. The computed model is shown in red.}
\end{figure}

We obtained photometry simultaneous to our spectroscopic HARPS observations
using EulerCam, the CCD imager at the 1.2m Euler-Swiss telescope, also located 
at La Silla Observatory. For each transit event, we extracted differential light curves
using relative aperture photometry, iteratively choosing a set of stable reference stars. 
Details on EulerCam and the related data reduction procedures can be found in \citet{Lendl12}. 
The individual time-series observations are listed in Table \ref{tab:phot}, together with a number of key properties and 
the light curves are shown in Figure \ref{fig:phot}.

\begin{table}
\centering   
\caption{\label{tab:phot}Overview of the simultaneous EulerCam photometric observations.}
\begin{tabular}{lllll} \hline \hline
date & filter &  $\beta_{r}$ & $\beta_{w}$ & $\mathrm{RMS_{5min}}$ [ppm] \T \\
\hline
11 Nov 2012 & r'-Gunn & 1.97  & 0.83 & 1184 \T \\
24 Oct 2017 & r'-Gunn & 1.80  & 0.65 & 703 \\
22 Nov 2017 & IC      & 1.34  & 1.01 & 860 \\
\hline
\end{tabular}
\end{table}

We jointly analysed the ensemble of photometric data using the differential-evolution MCMC code described in \citet{Lendl17b}, which makes use of the transit model of \citet{Mandel02} and the MC3 sampler \citep{Cubillos17}. To account for instrumental systematics, we tested a range of photometric baseline models for each light curve, selecting the optimal model via comparison of the Bayesian Information Criterion. For all three light curves, we found significant (Bayes factor $>$ 100) improvements when including a linear trend in stellar FWHM next to our minimal model of a linear trend in time. For the light curve of 24 October 2017, we found further improvements when including a second-order polynomial in coordinate shifts of the stellar PSF center.

To account for additional red noise, we rescaled photometric errors via the $\beta_{r}$ and 
$\beta_{w}$ factors as described in \citet{Gillon10a} and \citet{Winn08}. We assumed a quadratic 
limb-darkening law and used the routines of \citet{Espinoza15} to compute coefficients tailored to
WASP-76, the EulerCam detector efficiency and filter transmission. To account for the uncertainty 
affecting the limb-darkening coefficients, we assumed Gaussian prior distributions in our analysis, 
centered on the computed values, and having widths corresponding to the offsets incurred when altering the stellar parameters within the 1-$\sigma$ uncertainties given in \citet{Brown17}. 

Based on the observed light curves, we recomputed stellar and planetary parameters using the 
newly-available stellar radius measurement of WASP-76 from Gaia \citep{gaia18}. With a revised stellar radius of $1.969\pm0.035$~{\Rsolar} (compared to $1.7\pm0.03$~{\Rsolar}, \citealt{Brown17}), we find a $\sim$20\% larger planetary radius of $2.078_{-0.044}^{+0.036}$~{\Rjup} (compared to $1.73 \pm 0.03$~{\Rjup}). As already hinted at in \citet{Brown17}, we find a larger impact parameter compared to that stated in the discovery paper \citep{West16}. A full list of the derived stellar and planetary parameters is given in Table \ref{tab:para}.

\begin{table}[h]
\centering   
\caption{\label{tab:para}Planetary and stellar parameters inferred from EulerCam photometry.}
\begin{tabular}{ll} \hline \hline
\multicolumn{2}{l}{Jump parameters:} \T  \\
\hline
Mid-transit time,[BJD] - 2450000                \T & $  8080.62487\pm 0.00018                   $ \\
 $ R_{p}/R_{\ast} $                             \T & $  0.10824 \pm 0.00081                     $ \\
Impact parameter                                \T & $  0.23  _{-0.11}^{+0.080}                 $ \\
Transit duration [d]                            \T & $  0.15818   _{-0.00066}^{+0.00068}        $ \\
Period [d]                                      \T & $  1.80988145 _{-0.00000028}^{+0.00000020} $ \\
$ c_{1,\rm r}=2u_{1,\rm r}+u_{2,\rm r} $        \T & $  0.993  _{-0.040}^{+0.050}               $ \\
$ c_{2,\rm r}=u_{1,\rm r}-2u_{2,\rm r} $        \T & $  0.157 _{-0.056}^{+0.061}                $ \\
$ c_{1,\rm IC}=2u_{1,\rm IC}+u_{2,\rm IC} $     \T & $  0.703 _{-0.042}^{+0.051}                $ \\
$ c_{2,\rm IC}=u_{1,\rm IC} -2u_{2,\rm IC} $    \T & $  0.089 _{-0.063}^{+0.056}                $ \\
\hline                                             
\multicolumn{2}{l}{Deduced parameters:} \T                                                        \\
\hline                                             
Planetary radius, $ R_{p} $ [{\Rjup}]            \T & $ 2.078  _{-0.044}^{+0.036}               $ \\
Transit depth, $ \Delta F$ \T                    \T & $ 0.01171_{-0.00017}^{+0.00018}           $ \\
$ a/R_{\ast} $                                   \T & $ 4.08   _{-0.11}^{+0.02}                 $ \\
Orbital semi-major axis, $ a $ [au]              \T & $ 0.03675_{-0.00084}^{+0.00098}           $ \\
Stellar mean density, $ \rho_{\ast} $ [{\rhosun}] \T & $ 0.279  _{-0.022}^{+0.004}              $ \\
Stellar mass, $ M_{\ast} $ [{\Msolar}]           \T & $ 2.02   _{-0.14}^{+0.15}                 $ \\
$R_\ast$  [{\Rsolar}]   (retrieved)          \T & $ 1.968  _{-0.034}^{+0.036}               $ \\ 
$R_\ast$  [{\Rsolar}]     (GAIA)      \T & $ 1.969  _{-0.031}^{+0.035}               $ \\
Inclination [deg]                                \T & $ 86.72  _{-1.18}^{+1.72}                 $ \\
Eccentricity, $ e $ (fixed)                      \T & $ 0.0                                     $ \\
\multicolumn{2}{l}{Limb-darkening coefficients:} \T   \\
  $u_{1,\rm r} $      \T & $ 0.387 _{-0.027}^{+0.024}                $ \\
 $ u_{2,\rm r} $      \T & $ 0.235 _{-0.051}^{+0.031}                $ \\
 $ u_{1,\rm IC} $     \T & $ 0.272 _{-0.035}^{+0.017}                $ \\
 $ u_{2,\rm IC} $     \T & $ 0.178  \pm 0.043                        $ \\
 \hline
\end{tabular}
\end{table}

\section{Transmission spectroscopy of WASP-76b}
\label{sec:fulltrans}
\subsection{Transmission light curve}

\begin{figure*}[htb]
\resizebox{\textwidth}{!}{\includegraphics[trim=3.5cm 8.5cm 4.0cm 10cm]{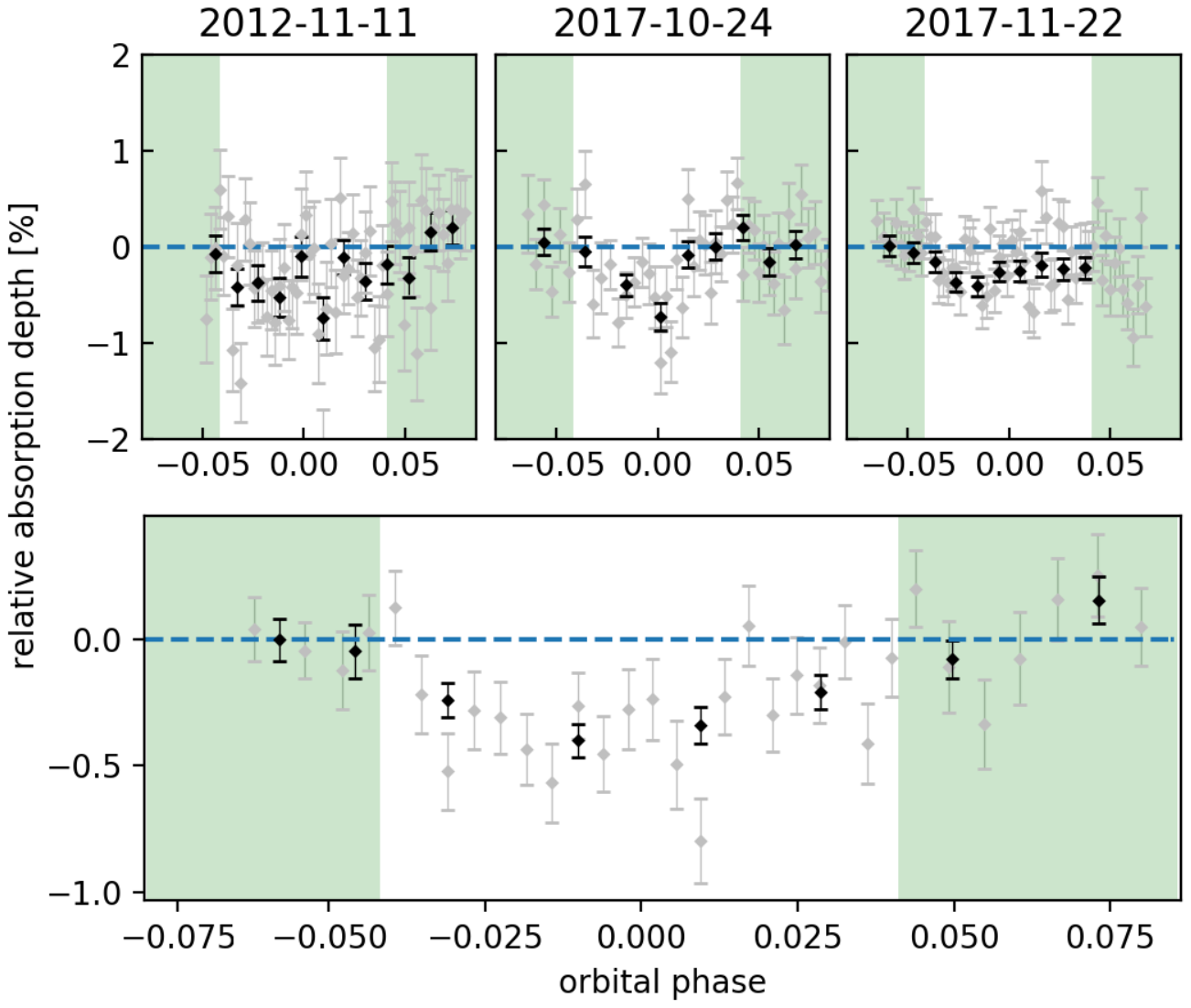}}
	\caption{HARPS relative excess absorption in the sodium doublet compared to the white light curve of WASP-76b evolving in time.  The upper three plots show the light curves for each night 2012-11-11, 2017-10-24 and 2017-11-22 from left to right. Note the different scales on the ordinate axis for the upper and lower panel and the influence of stratus-clouds in the out-of-transit exposures during the night of 2017-11-22. The grey data points show the relative absorption for each exposure, in black the data is binned by 5 spectra. The lower panel shows the light curve for all three nights combined, with the exposures averaged together with bins of 5 spectra in grey. The green background marks the exposures taken out-of-transit.}
	\label{fig:lc}
\end{figure*}

To properly depict the transit of the planet in front of its host star from the spectroscopic observations, we use the overall change of flux for the different exposures to construct the relative transmission light curve as a function of time \citep{Ch02,Sn08}. 
Figure \ref{fig:lc} depicts the transit depth as a function of time, which means that the wavelength dependence is lost. We derive the change of transit depth by comparing the relative flux at the center of the stellar sodium lines to the relative flux in reference passbands for each telluric corrected exposure. The reference passbands should remain unchanged during the planet's transit given that they only contain stellar lines. The passbands at line center, however change relative flux depending on whether they were taken in- and out-of-transit because they may contain the additional sodium absorption of the planet's atmosphere. Common reference passbands to use around the sodium doublet are bands on the blue (B) and red (R) side of the sodium doublet (window of $12$~\r{A}) of $12$\r{A} each in width \citep{Wy15}. The passbands in the sodium line centres (C$_{D_1}$ and C$_{D_2}$) are of $0.75$~\r{A} width. The relative flux ($\mathfrak{F}_{\mathrm{rel}}$) is calculated by averaging over the spectrum in the mentioned bands (represented by a bar over the quantity in Eq. \ref{eq:relflux}). Further discussion of the passbands can be found in section \ref{sec:transspec}.

The relative flux for each exposure is calculated with:

\begin{equation}
\label{eq:relflux}
\mathfrak{F}_{\mathrm{rel}}(t,\Delta \lambda)  = \frac{\overline{F(C_{D_1})}+\overline{F(C_{D_2})}}{\overline{F(B)}+\overline{F(R)}}
\end{equation}

All lightcurve points $\mathfrak{F}_{\mathrm{rel}}(t_{\mathrm{out}},\Delta \lambda)$ calculated with out-of-transit spectra are then used to normalise the out-of-transit light curve to zero. \cite{As13} showed that this method, using simple average differences, produces reliable results because differential stellar limb-darkening does not affect the measurements on narrow passbands as the ones selected in this work. The light curve obtained for the night of 2017-11-22 shows a visible slope in the post-transit absorption depth with time (and airmass), which has a significant effect on the transmission spectrum as discussed in section \ref{sec:transspec}. The observer confirmed the presence of stratus-clouds during these exposures. 

The night in question has ten exposures out-of-transit before the start of the transit which allows for the affected exposures to be disregarded from the analysis of the transmission spectrum. The light curves for each of the three observed nights and the overall light curve is shown in Fig. \ref{fig:lc}. 

The additional absorption depth during the planet transit can be seen in each of the three nights and even clearer when combining all three transits by comparing the in-transit data to the data taken out-of-transit (highlighted with a green background in Figure \ref{fig:lc}).

\subsection{Extraction of the transmission spectrum}
\label{sec:spec}

Following \cite{Br01} to calculate the spectral ratio, $\mathfrak{R}$, all e2ds spectra are corrected for telluric signatures and stacked together, in and out-of-transit, respectively, to form a master spectrum in-transit ($\mathcal{F}_{\mathrm{in}}(\lambda)$) and a master spectrum out-of-transit ($\mathcal{F}_{\mathrm{out}}(\lambda)$). The division of master-in by master-out spectrum gives the classical spectrum ratio, eliminating the stellar features and leaving only the planetary atmospheric signature $\mathfrak{R}= \mathcal{F}_{\mathrm{in}}(\lambda)/\mathcal{F}_{\mathrm{out}}(\lambda)$.
Ground-based observations do not allow for a direct calculation of the spectrum ratio due to flux changes with time. To reduce this effect we fitted a polynomial (order 3) to the continuum of the ratio $\delta = f(\lambda, t)/\mathcal{F}_{\mathrm{out}}(\lambda)$ and divided each exposure $f(\lambda, t)$ obtained at a given time $t$ by the fitted continuum. The fit to the ratio $\delta$ assures that the stellar lines do not influence the fit \citep{Al17}.

 The normalised spectra $ \tilde{f}(\lambda, t_{\mathrm{in}})$ and $\tilde{f}(\lambda, t_{\mathrm{out}})$ are then used to calculate the similarly self-normalised master-in $\tilde{\mathcal{F}}(\lambda, t_{\mathrm{in}})$ and master-out $\tilde{\mathcal{F}}(\lambda, t_{\mathrm{out}})$. 
We normalise all quantities on the in-transit spectra following:
\begin{equation}
\label{eq:norm}
 \tilde{f}(\lambda, t_{\mathrm{in}})=\frac{f(\lambda, t_{\mathrm{in}})}{f(\langle\lambda_{\mathrm{ref}}  \rangle, t_{\mathrm{in}})}
\end{equation}
However the spectral ratio $\tilde{f}/\tilde{\mathcal{F}}_{\mathrm{out}}$ does not consider changes in radial velocity of the planet since it is calculated in the stellar rest frame. As WASP-76b transits its star, the radial velocity changes between the maximum values of $-50$ km s$^{-1}$ and $+50$ km s$^{-1}$, which means that the planetary sodium signature shifts from the blue-shifted to the red-shifted part of the stellar sodium lines during the transit. In the wavelength range near the sodium doublet the shift is $\lesssim 1$\r{A} during the transit. We correct for this shift by transferring each calculated spectral ratio in the planet rest frame.

The end result of the new spectrum ratio is:

\begin{equation}
\mathfrak{\tilde{R}}(\lambda)= \sum\limits_{t \in \mathrm{inTransit}} \frac{\tilde{f}(\lambda, t)}{\tilde{\mathcal{F}}_{\mathrm{out}}(\lambda)}\bigg\rvert_{\mathrm{Planet RV shift}}
\end{equation}

\begin{figure*}[htb]
\resizebox{\textwidth}{!}{\includegraphics[trim=3.0cm 9.0cm 4.0cm 9.5cm]{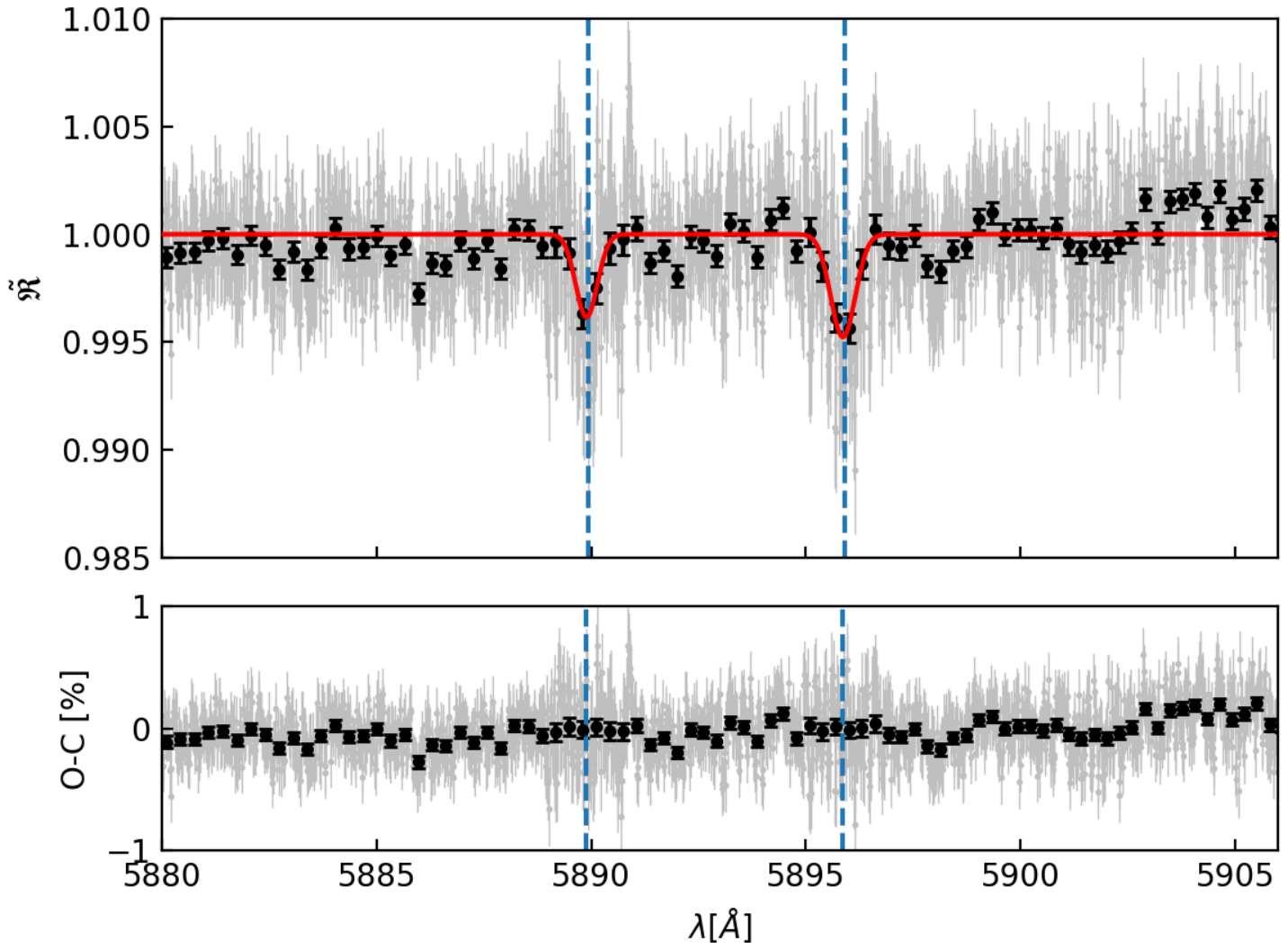}}
	\caption{HARPS transmission spectrum of WASP-76b centred around the sodium doublet in the planetary rest frame. Upper panel: Light grey - Full transmission spectrum for all nights combined in full resolution. Black - Transmission spectrum binned by x20. The data has been corrected for tellurics and cosmics and shifted to the planetary rest frame. The absorption from the two Na I D lines from the planetary atmosphere can be seen clearly and the rest frame transition wavelengths are marked with blue dashed lines. 
	 Red - Gaussian fits to each Na I line. We measure line contrasts of $0.373\pm0.091\%$ (D2) and $0.508\pm0.083\%$ (D1) and FWHMs of  $0.619\pm0.174$\r{A} and $0.680\pm0.128$\r{A} respectively. Lower panel: Residuals of the Gaussian fit. }
	\label{fig:transspectrum}
\end{figure*}

A more detailed description of the calculation of the  spectrum ratio can be found in \cite{Wy15}. The transit spectrum of WASP-76b taking into account all three nights is plotted in the middle panel of Fig. \ref{fig:transspectrum}.

\subsection{Binned atmospheric absorption depth}
\label{sec:transspec}

The upper panel of Fig. \ref{fig:transspectrum} contains the transit spectrum of WASP-76b in the planetary rest frame and shows a double peaked feature, one peak for each component of the sodium doublet.
\begin{figure}
\resizebox{\columnwidth}{!}{\includegraphics[trim=3.5cm 9.2cm 3.0cm 10.5cm]{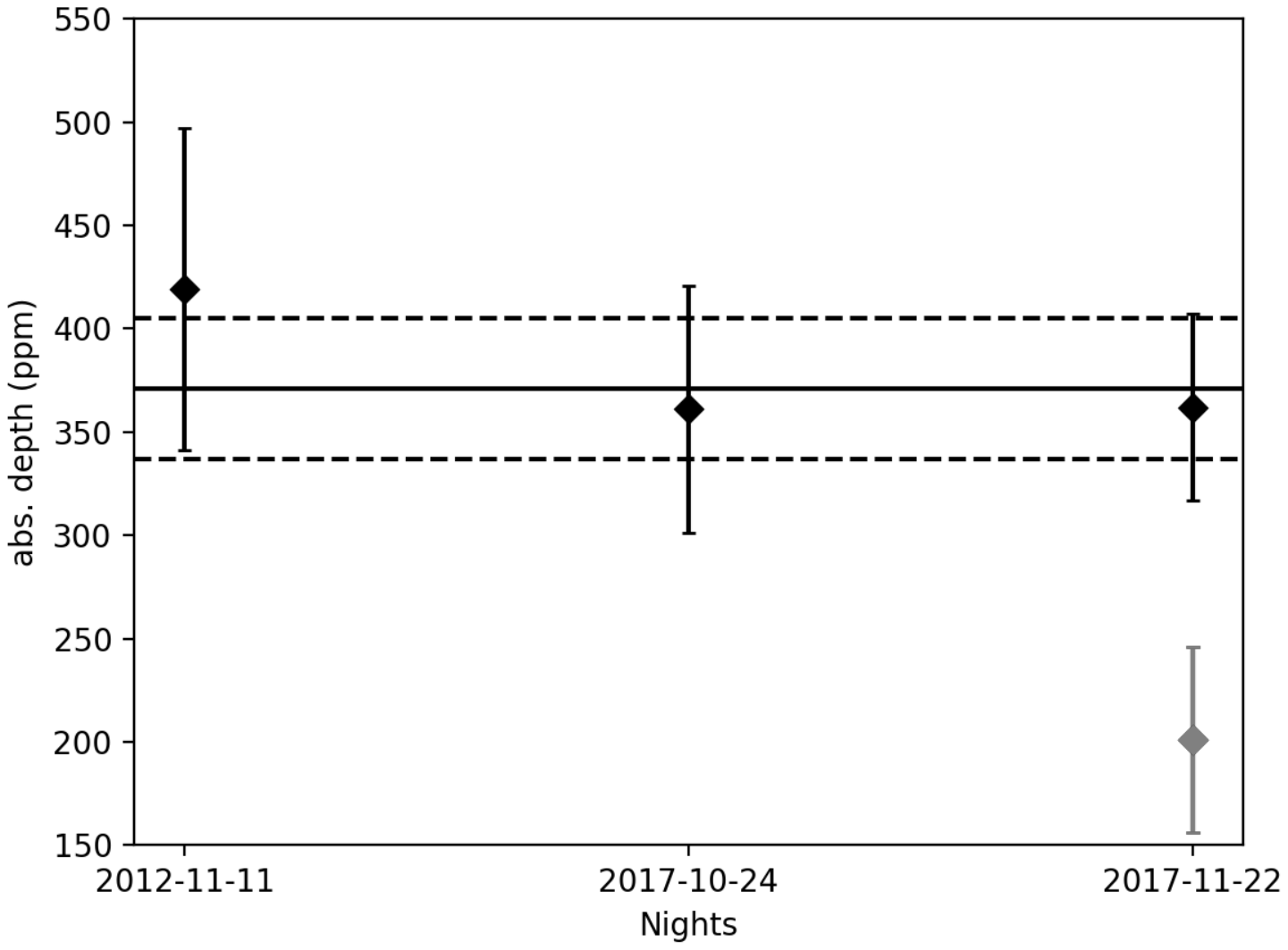}}
	\caption{The absorption depth of the three nights calculated from the transmission spectrum of the sodium doublet plotted with the value for all nights together highlighted as a black horizontal line. The dashed lines mark one sigma around the main absorption depth. The data point in gray shows the value calculated for Night 3 when the cloud contaminated exposures are not discarded. All values were taken from Tables \ref{table:absorbnightly} and \ref{table:absorb}.}
	\label{fig:alldepth}
\end{figure}

The features are close to the wavelengths of the sodium doublet lines (D$_1$ and D$_2$) at wavelengths $5895.924$~\r{A} and $5889.951$~\r{A}, respectively. The systemic velocity of WASP-76 is $\gamma = -1.07~\kms$, which shifts these features to  $5895.903$~\r{A} (D$_1$) and $5889.930$~\r{A} (D$_2$) in the solar system barycentric reference frame. 
The telluric sodium lines should be located around the BERV values of $-12.6$,$-3.6$ and $-17.5~\kms$ for each of the nights and have no impact on the planetary signal. 

The relative absorption depth can be calculated by averaging the flux in the sodium doublet line cores and comparing it to the average flux in specified reference bands in the continuum. Contrary to \cite{Sn08}, who chose adjacent control passbands in the red (R) and blue (B) of the sodium doublet, we employ the approach of \cite{Ch02} with absolute reference bands outside of the sodium doublet. This approach is justified given that the high-resolution data resolves the doublet. The two reference bands were chosen as follows: $[5874.89-5886.89]$~\r{A} for the red band (R) and $[5898.89-5910.89]$~\r{A}  for the blue band (B), corresponding to 12~\r{A}  each.

The central passband (C) containing the sodium signal is split in two centered passbands, one for each line of the doublet. The width of the passband is varied from containing only the line core to a broader passband reaching from the red to the blue reference band. We chose the following central passband widths in line with \cite{Wy15,Wy17}: $2\times0.75$~\r{A}, $2\times1.5$~\r{A}, $2\times3$~\r{A}, $2\times6$~\r{A}.

The relative depth caused by the exoplanetary atmospheric absorption is then obtained via the flux difference between the central and the reference passbands via:

\begin{equation}
\delta(\Delta\lambda)=\overline{\mathfrak{R}(C)}-\frac{\overline{\mathfrak{R}(B)}+\overline{\mathfrak{R}(R)}}{2}
\end{equation}

The detection levels for the different passbands can be found in Table \ref{table:absorb} and show that the detection of planetary sodium in WASP-76b is independent of the passband choice. Nonetheless, the passband restricted around the line center ($2\times0.75$~\r{A}) shows the strongest sodium feature, as expected, with a detection level of $10.75 \sigma$. The detection in this passband (see Table \ref{table:absorb}) corresponds to an atmospheric height of $28\,800\pm 2\,600$ km ($0.19\pm0.02\mathrm{R_p}$) \citep{Pi17}.

\begin{table}
\caption{Relative depth and detection level of atmospheric sodium on WASP-76b observed with HARPS for different central passbands\tablefootmark{a}.}
\label{table:absorb}
\centering
\begin{tabular}{c c c }
\hline
\hline
central passband (C)& abs. depth	(\%)& $\sigma$	\\
\hline
$2\times6.00$ \r{A}& $0.108\pm0.012$&$9.15$\\
$2\times3.00$ \r{A}&$0.152\pm0.017$&$9.05$\\
$2\times1.50$ \r{A}&$0.223\pm0.025$&$9.10$\\
$2\times0.75$ \r{A}&$0.371\pm0.034$&$10.75$\\
\hline
\end{tabular}
\tablefoot{\tablefoottext{a}{Excluding all data after the planet's transit for night 3 (2017-11-22) due to clouds. The baseline of the specified night was created from the data obtained before the transit.}}
\end{table}

To establish the repeatability of the detection, we compare the relative depth for the $2\times0.75$ \r{A} central passband for all nights separately (see Table \ref{table:absorbnightly} and Figure \ref{fig:alldepth}). The values for all three nights lie within $\pm1\sigma$, which means the detection of sodium in the planetary atmosphere of WASP-76b is repeatable. The out-of-transit exposures affected by the drop in flux due to clouds during the last hours of the third night (2017-11-22) (see the data in the upper right corner of Fig. \ref{fig:lc}) change the detection level significantly. As stated before, 10 unaffected exposures were taken before the transit, which can be used to confidently establish a baseline without the contaminated exposures.

\begin{table}
\caption{Relative depth and detection level of atmospheric sodium on WASP-76b observed with HARPS for all nights separately in the $2\times0.75$ \r{A} central passband.}
\label{table:absorbnightly}
\centering
\begin{tabular}{l c c c }
\hline
\hline
&Date& abs. depth	(\%)& $\sigma$	\\
\hline
Night 1&2012-11-11& $0.419\pm0.078$&$5.46$\\
Night 2&2017-10-24&$0.361\pm0.060$&$6.06$\\
Night 3\tablefootmark{a}&2017-11-22&$0.362\pm0.045$&$8.02$\\
\hline
Night 3\tablefootmark{b}&2017-11-22&$0.201\pm0.045$&$4.51$\\
\hline
\end{tabular}
\tablefoot{\tablefoottext{a}{Excluding all data after the planet's transit for night 3 (2017-11-22). The baseline of the specified night was created from the data obtained before the transit.}\tablefoottext{b}{Including all datapoints obtained during night 3.}}
\end{table}

\subsection{Center-to-Limb variation and Rossiter-McLaughlin effect}

Additional effects that could impact this analysis are Center-to-Limb Variation (CLV) and the Rossiter-McLaughlin effect (RM). As the stellar spectra obtained by HARPS are integrated over the stellar disk, both effects have the potential to significantly alter the transmission spectrum \citep{Lo15}.

 To analyse the impact of the RM effect on our analysis, we built average transmission spectra in the star and planet rest frame (see equation \ref{eq:norm} and section \ref{sec:spec}) and calculated the CCF with the G2 mask on the red detector (700 lines). This allows to measure the average change in line shape imprinted by the Rossiter-McLaughlin effect on the transmission spectrum. This method is comparable to the "transmission ccf" method presented in \cite{Wy17} (see also \cite{Bo18}). The feature expected from the RM at 0 \kms is not detected beyond the noise level of 100 ppm. The small noise level is due to the peculiar characteristics of WASP-76, which is a slow rotator ($\vsini < 2.3 \kms$) and has its planet, WASP-76b, in a polar orbit \citep{Brown17}. As a conclusion, the planet always masks an area of the star with almost zero velocity during the whole transit. Since the 100 ppm value is about 4 times smaller than our detection, we conclude that the RM effect cannot explain our detection and is subsequently neglected.\\
 
For WASP-76, a F7 star, we  assume that CLV effects can be disregarded given that CLV effects are less pronounced for earlier type stars \citep{Ko15, Cz15, Ya17}. According to the theoretical modelling of \cite{Cz15}, the  impact of CLV effects on our data should be negligible: these authors highlight a star similar to WASP-76 (F7, $T_{\mathrm{eff}} = 6\,120K$) and CLV effects remain below the threshold of $200$--$500$ ppm). Given that our light curve depth is about 0.005 (factor of 10 stronger) we do not take these effects into account during the following analysis.

\subsection{Systematic effects}

\begin{figure*}[htb]
\resizebox{\textwidth}{!}{\includegraphics[trim=1.5cm 6.5cm 0.75cm 6.5cm]{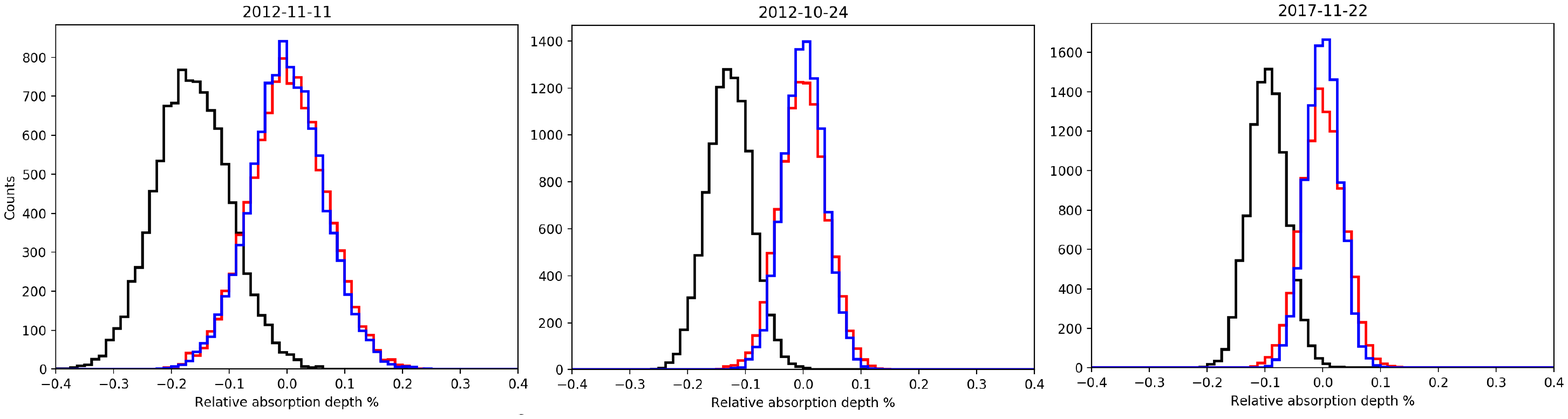}}
	\caption{Distributions of the empirical Monte-Carlo analysis for the $12$ \r{A} passband. The results for the transmission spectrum method are shown. As expected the 'in-in' (red) and 'out-out' (blue) distributions are centred around $0$ (no planetary detection) and the randomised 'in-out' distribution shows a detection (black). In each night a different number of spectra was taken, which means that each randomisation will give a different number of counts, which results in a different scaling on the the y-axis. This has no influence on the overall result.}
	\label{fig:random}
\end{figure*}

To calculate the error on the relative absorption depth ($\delta(\lambda)$) we propagated the error from the measured spectra through our analysis. The initial error is estimated as random photon noise obeying Poisson statistics. However, systematic errors and spurious signals cannot be ruled out. To further investigate their impact on our results we employ empirical Monte-Carlo (EMC) or bootstrapping methods following the approach in \cite{Re08}. 
The main goal is to verify that our signal is indeed produced by a planetary atmosphere and not by a  random arrangement of the data,  however statistically unlikely.
To this end, we create different 'scenarios' for comparison. The first scenario represents the case of an atmospheric detection: we select at random a sub-sample of spectra from the in-transit spectra and also at random the same number of out-of-transit spectra. These spectra are then used to calculate a randomised master-out spectrum ($\tilde{\mathcal{F}}_{\mathrm{out}}(\lambda)$) and subsequently $\delta(\lambda)$. This scenario should yield a detection and will be called the 'in-out' scenario.
Additionally a sub-sample containing only spectra from in-transit exposures was taken similarly and then split into two data sets of equal size. One half is labeled the 'virtual in-transit' sample and the other the 'virtual out-of-transit' sample. 

This scenario is then called the 'in-in' scenario. In the same fashion, 'out-out' scenarios are created from the out-of-transit data. We created $10,000$ 'in-out', 'in-in' and 'out-out' scenarios each, and run our transmission spectroscopy analysis with them as described in section \ref{sec:transspec}. The resulting relative absorption depth distributions are plotted in Fig. \ref{fig:random} for the three observed nights. The red and blue distributions show the relative absorption depth for the 'in-in' and 'out-out' scenarios respectively. A Gaussian fit to the distributions shows that both are centered at $0.0\%$ as expected. The 'in-out' distribution (in black) shows a relative absorption distribution centered around $-0.1\%$. 
This rules out a spurious detection of a planetary atmosphere as a statistical anomaly.

\subsection{Broadening of the sodium feature}
\begin{figure*}
\resizebox{\textwidth}{!}{\includegraphics[trim=3.0cm 9cm 3.5cm 10.5cm]{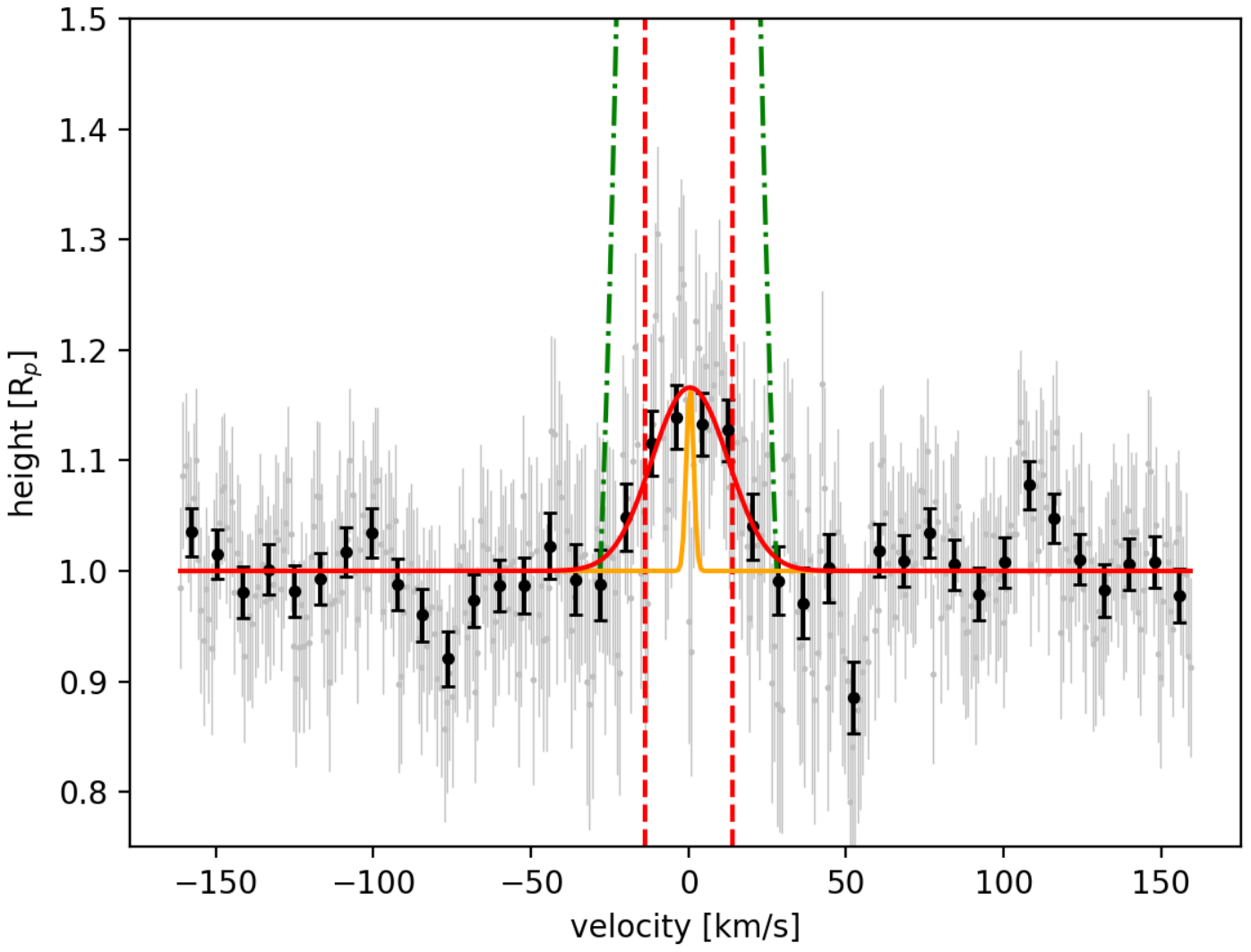}}
	\caption{The co-added lines of the HARPS transmission spectrum sodium doublet as a function of velocity. The line center was set to zero. The data is shown in grey and binned x10 in black. Gaussian fit derived as in Section \ref{sec:spec} is shown in red with its FWHM marked by the red dashed lines. The HARPS instrumental line spread function (FWHM = $2.7 \kms$) is shown in orange and the escape velocity as green dotted-dashed lines. The FWHM velocity does not exceed the escape velocity at any point, making the atmospheric escape of sodium unlikely. The best Gaussian fit to the data is significantly wider than the instrumental response, but the corresponding velocity is well below the escape velocity of WASP-76b. Given that the conversion of relative depth to height depends on the choice of height at the continuum level (here: WASP-76b white light radius), the height is not an absolute height but an equivalence height \citep{Pi17}.}
	\label{fig:coadded}
\end{figure*}

Both \ion{Na}{i} lines are substantially broader than the instrumental line spread function of HARPS (2.7~\kms) as shown in Fig.~\ref{fig:coadded} where the two lines of the doublet are co-added. We first fit a Gaussian profile to measure the FWHM and the amplitude of the sodium lines. We find that a Gaussian profile provides a good fit to the data ($\chi^2$ of 126.48 with 194 degrees of freedom). 

First, we note that the depth of the D$_1$ and D$_2$ lines, as measured from the contrasts of the Gaussian fit in Fig.~\ref{fig:transspectrum}, is compatible to within 1--2$\sigma$, as expected for the depths of the \ion{Na}{i} lines seen in absorption in a transit transmission spectrum. 

The resulting FWHM is  $27.6\pm2.8~\kms$ which is $10.2 \times$ broader than the resolution element of the spectrograph. The line is substantially broader than what we would expect from pure thermal broadening or a super-solar abundance of sodium. We verified this using the $^\pi\eta$ transit transmission spectroscopy code \citep{Pi17}. In contrast with previous work \citep{Wy15}, we were not able to adjust the line profile either by increasing the temperature (up to $\sim17,000$~K ) or varying the mixing ratio of sodium (100$\times$ solar abundance down to $10^{-6} \times$ solar abundance). The \ion{Na}{i} lines in WASP-76b have a similar depth compared to HD~189733b (Wyttenbach et al. 2015), but are $\sim24~\%$ broader. We therefore attribute the broadening to missing physics in the hydrostatic atmospheric model used for the transit spectroscopy calculation.

One possibility is that the line broadening is of kinematic origin. A high-altitude atmospheric circulation around the planet (atmospheric super-rotation, typically not accounted for in  1-D hydrostatic atmospheric models) could produce a Doppler broadening:  sodium atoms moving towards and away from the observer at the evening (West) and morning (East) terminator, would respectively broaden the blue and red wing of the line. 

Our data suggest that most sodium atoms have projected velocities below the escape velocity of $28$~\kms\, calculated at a planetary radius of $2.08$~R$_\mathrm{Jup}$ (Fig.~\ref{fig:coadded}). Assuming a Boltzmann distribution for the particle velocities, we calculate that only $\sim0.1\%$ of all particles have velocities greater or equal to the escape velocity. While these few sodium atoms in the tail of the velocity distribution may escape the planet, Fig.~\ref{fig:coadded} shows that the bulk have velocities well within the escape velocity; therefore the absorption signature is mainly caused by atoms gravitationally bound to the planet.
Because we only measure the projected velocities of the sodium atoms, the fraction of high velocity sodium atoms might be higher, but this effect is unlikely to change our conclusion.

To better constrain atmospheric escape, follow-up studies need to trace layers above the ones probed by sodium, using lighter gases such as hydrogen and helium. Additionally, WASP-76b is too far away from Earth to measure UV transit absorption signal in the stellar H I Lyman-$\alpha$ emission line (which is entirely absorbed by the interstellar medium), but there are good prospects to observe escaping helium through high-resolution, near-infrared transit observations \citep{Al18,No18,Sa18}.

\section{Conclusion}
\label{sec:conclusion}

We have analysed 160 spectra of WASP-76, with 107 spectra while WASP-76b was in-transit, during three nights of observations with HARPS. This led to the first detection of sodium in the planet atmosphere from the transmission spectrum ($0.371 \pm 0.034 ~\%$; $10.75~ \sigma$ in the $0.75$~\r{A} passband) in an ultra-hot Jupiter. The result was reproducible in each of the three obtained nights. We were able to rule out spurious signals from data artifacts via EMC analysis by testing for a detection from random exposures containing only in- of out-of-transit exposures, respectively.
We co-added the sodium doublet and retrieved a FWHM of $27.6\pm2.8~\kms$ for the fitted Gaussian. The significant line broadening could also be a result of rapid, large-scale circulation of hot atoms in the upper atmosphere of the planet, leading to Doppler broadening. This hints towards upper atmospheric super-rotation in an ultra-hot gas giant. The bulk of the population of hot neutral atoms do not escape the planet.

The HEARTS survey is intended as a pathfinder for a survey using the ESPRESSO spectrograph, which started operations in October 2018 at the 8-meter-class Very Large Telescope. 
These new observations at high spectral resolution will substantially increase our understanding of exoplanetary atmospheres in different irradiation conditions and will shed light on atmospheric conditions such as winds and ionisation state.


\begin{acknowledgements}
This project has received funding from the European Research Council (ERC) under the European Union's Horizon 2020 research and innovation programme (project {\sc Four Aces}; grant agreement No. 724427).
This work has been carried out within the frame of the National Centre for Competence in Research `PlanetS' supported by the Swiss National Science Foundation (SNSF). A. W., R. A. acknowledge the financial support of the SNSF (A. W.: Nr. P2GEP2\_178191). N. A.-D. acknowledges support from FONDECYT Nr. 3180063. We are grateful to J. M. D\'esert and V. Panwar for allowing us to use their computational resources to run the $^\pi \eta{}$ code and thank L. DosSantos for his helpful comments. Additionally, we would like to highlight the contribution of the anonymous referee.
\end{acknowledgements}
%

\bibliographystyle{aa} 
\bibliography{HEARTSIIWASP76b} 
\end{document}